# Trapped magnetic field in a superconducting disk magnetized by uniform field

**C Li, J J Wang, C Y He, L F Meng, R S Han, Z X Gao**

Department of Physics, Key Laboratory for Artificial Microstructure and Mesoscopic Physics, Peking University, Beijing 100871, P. R. China

**Abstract**

The distribution of the current density $J(\rho, z)$ and the profile of the trapped magnetic field of a superconducting disk magnetized by uniform field are calculated from first principles. The effect of the superconducting parameters is taken into account by assuming the voltage-current law $\boldsymbol{E} = \boldsymbol{E}_c (J/J_c)^n$ and the material law $\boldsymbol{B} = \mu_0 \boldsymbol{H}$. The sample volume, the critical current density and the flux creep exponent dominate the strength of the trapped magnetic field. The flux creep exponent determines the decay rate of the trapped magnetic field. The aspect ratio $b/a$ of the sample influences the shape of the trapped magnetic flux lines. We conclude that higher trapped magnetic field cannot always be obtained by increasing the applied magnetic field.

PACS number(s): 74.60.2w, 74.25.Ha, 74.25.Ld

## 1. Introduction

A large light rare earth (LRE) - $Ba_2 Cu_3 O_{7-x}$ (e.g., Nd, Sm, Eu, or Gd) superconducting bulk magnet (SBM) is believed to trap very high magnetic field more than 5 T at 77 K. The melt-textured-grown (MTG) superconducting quasi-permanent magnet has received intense attention from the point of view of practical applications because of its strong pinning force at 77 K [1-3]. Magnetic levitation train, flywheels and magnetic bearings have been developed by utilizing the repulsive force against the magnetic field produced by a permanent magnet (PM). The superconductor with trapped magnetic field is used as a permanent magnet in a superconducting motor, magnetic separation and a magnetic field generator capable of producing fields higher than 2 T [4].

To use the MTG bulk superconductor as quasi-permanent magnet, it is necessary to magnetize it beforehand. The magnetization by uniform field is the most effective method to extract the potential of this material. The trapped magnetic field in the SBM is the key property for its applications. In this paper we have calculated the trapped magnetic field of a superconducting disk after magnetizing it by uniform field from first principles.

In previous calculations, the current density and the magnetic field of a finite cylinder in an axial magnetic field are calculated from first principles [5-7]. In those cases they discussed the process that the magnetic flux penetrates into the finite cylinder. In our paper, we focus on the trapped magnetic flux escaping from the disk.

In practice the superconducting disk can be cooled below $T_c$ in an axial applied uniform magnetic field $B_a$, then the applied magnetic field is switched off, and the magnetic flux escapes from the disk. This is called the escape process. The calculation of current density in the superconducting disk will be divided into two steps with different applied magnetic fields. First, the surface screening current density $J_1'$ is induced just after switching off the axial applied uniform magnetic field such that $\dot{B}_a \neq 0$. Second, the current density $J_2'$ diffuses into the disk center after $\dot{B}_a = 0$ and $B = 0$. The result obtained in the first step is taken as the initial current density in the second step. It is difficult to calculate the above current densities $J_1'$ and $J_2'$ in a superconducting disk directly. Instead of direct calculation, we suggest a method that shuns such a difficulty.

We consider another process that the magnetic flux penetrates into the same superconducting disk. After the superconducting disk is cooled in zero magnetic field below its critical temperature, we start at time $t=0$ with $B_a = 0$ and then switch on an axial applied uniform magnetic field such that $\dot{B}_a \neq 0$. The surface screening current $J_1$ is induced. As soon as $\dot{B}_a = 0$ and $B_a = \text{const}$, the superconducting disk is in the axial applied uniform magnetic field. During the penetration of the magnetic flux into the disk, the current density $J_2$ generates. This is called the penetration process.

In the above escape and penetration processes, the disk, the uniform magnetic field and the flux motion are the same, so the gradient of magnetic field has the same value but in opposite direction, that is $J_1 = -J_1'$ and $J_2 = -J_2'$. The $J_1$ and $J_2$ can be calculated from the equation of motion for current density in a superconducting disk.

## 2. Modeling

### A. Basic equation for current density

We use the method presented by E. H. Brandt [5-7] to calculate the current density in a superconducting disk. The key issue is to find an equation of motion for the current density $J(\mathbf{r},t)$ inside the disk. Using the Maxwell equation $\mathbf{J} = \nabla \times \mathbf{H}$, and considering $\mathbf{B} = \mu_0 \mathbf{H}$, $\nabla \cdot \mathbf{A}_J = 0$ and $\nabla \times \mathbf{B}_a = 0$ (assuming no current sources i.e. no contacts), we obtain the equation of motion for the current density: $\mu_0 \mathbf{J} = \nabla \times \mathbf{B} = \nabla \times \nabla \times \mathbf{A}_J = -\nabla^2 \mathbf{A}_J$, where $\nabla \times \mathbf{B} = \nabla \times (\mathbf{B}_a + \mathbf{B}_J) = \nabla \times \mathbf{B}_J$ and $\nabla \times \nabla \times \mathbf{A}_J = \nabla(\nabla \cdot \mathbf{A}_J) - \nabla^2 \mathbf{A}_J$. As usual, the displacement current, which contributes only at very high frequencies, is disregarded in this "eddy-current approximation". We describe the superconductor by the material law $\mathbf{B} = \mu_0 \mathbf{H}$ and the voltage-current law $\mathbf{E} = E_c (J/J_c)^n$. $\mathbf{B} = \mu_0 \mathbf{H}$ is a good approximation when the flux density $B$ and the critical sheet current $J_c b$ are larger than the lower critical field $B_{c1}$ everywhere inside the superconducting disk [6,7]. This requirement is often satisfied in the magnetic levitation measurement. The voltage-current law $\mathbf{E} = E_c (J/J_c)^n$ is actually a result of the current dependence of the activation energy $U(J) = U_0 \log(J_c/J)$ by using the Arrhenius law $E = B\upsilon = B\upsilon_0 \exp[-U(J)/k_B T]$. So we obtain the parameters $E_c = B\upsilon_0$, $n = U_0/k_B T$ and $\upsilon = \upsilon_0 \exp(-U/k_B T)$, which is the effective vortex velocity. $\sigma(=n-1)$ is the flux creep exponent.

Because of the axial symmetry the current density $\mathbf{J}$, electric field $\mathbf{E}$, and vector potential $\mathbf{A}$ (defined by $\nabla \times \mathbf{A} = \mathbf{B}$) have only one component pointing along the azimuthal direction $\hat{\phi}$; thus $\mathbf{J} = J(\rho,z)\hat{\phi}$, $\mathbf{E} = E(\rho,z)\hat{\phi}$, and $\mathbf{A} = A(\rho,z)\hat{\phi}$. Since $\mathbf{B} = \mu_0 \mathbf{H}$, we have $\mu_0 J = -\nabla^2 A_J$, where $A_J = A - A_\phi$ is the vector potential generated by the current density in the disk, and $A_\phi$ is the vector potential of applied magnetic field. The solution of this Poisson equation in cylindrical geometry is

$$A(\rho, z) = -\mu_0 \int_0^a d\rho' \int_{-b}^{b} dz' Q(\mathbf{r},\mathbf{r}') J(\mathbf{r}') + A_\phi \quad (1)$$

with $\mathbf{r} = (\rho, z)$ and $\mathbf{r}' = (\rho', z')$. The integral kernel

$$Q_{cyl}(\mathbf{r},\mathbf{r}') = f(\rho, \rho', z - z') \quad (2)$$

with

$$f(\rho,\rho',z-z') = \int_0^\pi \frac{d\phi}{2\pi} \frac{-\rho'\cos\phi}{\left[(z-z')^2 + \rho^2 + \rho'^2 - 2\rho\rho'\cos\phi\right]^{1/2}}$$

$$= \frac{-1}{\pi k}\sqrt{\frac{\rho'}{\rho}}\left[\left(1-\frac{1}{2}k^2\right)K(k^2) - E(k^2)\right] \quad (3)$$

where

$$k^2 = \frac{4\rho\rho'}{(\rho+\rho')^2 + (z-z')^2} \quad . \quad (4)$$

$K$ and $E$ are the complete elliptic integrals of the first and second kind, respectively.

To obtain the desired equation of motion for $J(\rho,z,t)$, we express the induction law $\nabla\times\mathbf{E} = -\dot{\mathbf{B}} = -\nabla\times\dot{\mathbf{A}}$ in the form $\mathbf{E} = -\dot{\mathbf{A}}$. The gauge of $\mathbf{A}$ $(\nabla\cdot\mathbf{A}=0)$, to which an arbitrary curl-free vector field may be added, presents no problem in this simple geometry. Knowing the material law $\mathbf{E} = \mathbf{E}_c(J/J_c)^n$, we obtain $\dot{\mathbf{A}} = -\mathbf{E}(J)$. This relation between $\dot{\mathbf{A}}$ and $J$ allows us to eliminate either $A$ or $J$ from Eq. (1). After eliminating $A$, we obtain

$$E[J(\mathbf{r},t)] = \mu_0 \int_S d^2r' Q_{cyl}(\mathbf{r},\mathbf{r}')\dot{J}(\mathbf{r}',t) - \dot{A}_\phi(\rho',z') \quad (5)$$

This implicit equation for the current density $J(\mathbf{r},t)$ contains the time derivative $\dot{J}$ under the integral sign. In the general case of nonlinear $E(J)$, the time integration of Eq. (5) has to be performed numerically. For this purpose, the time derivative should be moved out from the integral to obtain $\dot{J}$ as an explicit functional of $J$. This inversion may be achieved by tabulating the kernel $Q_{cyl}^{-1}(\mathbf{r},\mathbf{r}')$ on a 2D grid $\mathbf{r}_i$, $\mathbf{r}_j$ and then inverting the matrix $Q_{ij}$ to obtain $Q_{ij}^{-1}$, which is the tabulated reciprocal kernel $Q_{cyl}^{-1}(\mathbf{r},\mathbf{r}')$. The equation of motion for the azimuthal current density $J(\rho,z,t)$ then reads

$$\dot{J}(\mathbf{r},t) = \mu_0^{-1}\int_0^a d\rho' \int_{-b}^b dz' Q_{cyl}^{-1}(\mathbf{r},\mathbf{r}')\left[E(J) + \dot{A}_\phi(\rho',z')\right] \quad (6)$$

where $Q^{-1}$ is the reciprocal kernel, which is defined by

$$\int_0^a d\rho' \int_{-b}^b dz' Q^{-1}(\mathbf{r},\mathbf{r}')Q(\mathbf{r}',\mathbf{r}'') = \delta(\mathbf{r}-\mathbf{r}'') \quad (7)$$

**B. Current density in SBM**

For an axial uniform applied magnetic induction $\mathbf{B}_a = B_a \hat{z}$, we choose the vector potential $A_\phi = -\frac{\rho}{2} B_a$. When we start at time $t = 0$ with $B_a = 0$ and $J = 0$ and then switch on the applied field such that $\dot{B}_a \neq 0$, the surface screening current is easily induced. Immediately after that, at time $t = +\varepsilon$, the magnetic induction and current density inside the disk are still zero since they need some time to diffuse into the conducting material. Therefore, at $t = +\varepsilon$ the electric field $E(J)$ is also zero and thus the first term in Eq. (6) vanishes. What remains is the last term, which should be the time derivative of the surface screening current $J_1$. This surface screening current is thus

$$J_1(\mathbf{r},t) = -H_a(t) \int_0^a d\rho' \int_{-b}^b dz' Q_{cyl}^{-1}(\mathbf{r},\mathbf{r}') \frac{\rho}{2} \tag{8}$$

The surface screening current lasts only very short time. As soon as $\dot{B}_a = 0$ and $B_a = \text{const}$, the motion for the current density in the disk must be described by

$$\dot{J}_2(\mathbf{r},t) = \mu_0^{-1} \int_0^a d\rho' \int_{-b}^b dz' Q_{cyl}^{-1}(\mathbf{r},\mathbf{r}') \left[ E_2(J_2) \right] \tag{9}$$

Eq. (9) is easily time integrated by starting with $J_2(\rho, z, t_{02}) = J_1$ and then putting $J_2(\rho, z, t = t + dt) = J_2(\rho, z, t) + \dot{J}_2(\rho, z, t) dt$.

**C. Trapped magnetic field**

As soon as the current density $J_2(\rho, z, t)$ is obtained, according to $J_2' = -J_2$, the vector potential induced by the current density can be derived by

$$A_J'(\rho, z) = -\mu_0 \int_0^a d\rho' \int_{-b}^b dz' Q(\mathbf{r},\mathbf{r}') J_2'(\mathbf{r}') \tag{10}$$

The trapped magnetic field in the cylindrical system can be readily calculated. The axial trapped magnetic field is

$$B_z = \frac{1}{\rho} \frac{\partial(\rho A_J')}{\partial \rho}, \tag{11}$$

and the radial trapped magnetic field is

$$B_\rho = -\frac{\partial A_J'}{\partial z}. \tag{12}$$

**3. Results and discussions**

Now we consider the value of the applied magnetic field $H_a (= B_a / \mu_0)$ in field cooling. For short cylinders in the Bean limit, we have an explicit expression for the field of full penetration $H_p$, i.e. the value of the applied magnetic field $H_a$ at which the penetrating flux and current fronts have reached the specimen center [6]. At $H_a > H_p$, the current density in the Bean model does not change anymore, and the trapped magnetic field created by the current density will not change either. For cylinders with arbitrary aspect ratio $b/a$, the field of full penetration is

$$H_p = J_c b \ln\left[\frac{a}{b} + \left(1 + \frac{a^2}{b^2}\right)^{1/2}\right] \qquad (13)$$

For $b/a = 1$, $H_p = J_c b \ln(1+\sqrt{2})$ and $B_p = \mu_0 J_c b \ln(1+\sqrt{2})$. Typically $a = 0.025$ m and $J_c = 2\times 10^4$ A/cm$^2$, then $B_p = 5.538$ T. The typical magnetic field applied to magnetize the melt-textured-grown YBCO disk is 1~2 T, so it is reasonable to take $H_a/H_p = 0.3$. We take $E_c = \mu_0 = J_c = 1$, $\omega = 0.1$, $\sigma = 20$, $b/a = 1$, and $H_a/H_p = 0.3$ in the following calculation, unless special declaration.

A. Surface screening current $J_1$

Fig. 1 shows the profile of the surface screening current density $J_1(\rho, z)$ calculated by Eq. (8). The thickness of this current-carrying layer depends on the choice of the computational grid in the disk. The layer thickness may be reduced, and the precision of the computed surface screening current can be enhanced, by choosing a non-equidistant grid, which is denser near the disk surface. An appropriate choice of such a non-equidistant grid $\mathbf{r}_i = (\rho_i, z_i)$ is obtained by taking $\rho = \rho(u) = \frac{1}{2}(3u - u^3)a$, and then tabulating $u = 0, \cdots, 1$ on equidistant grids $u_k = (k-\frac{1}{2})/N_\rho$ $(k = 1, \cdots, N_\rho)$. We divide $z$ into $N_z$ parts in the similar way, yielding a 2D grid of $N = N_\rho N_z$. The weights at different points should be considered. In this calculation we take $N \approx 400$.

A very interesting feature shown in Fig. 1(a) is that the screening current density in the side surface of the disk is much higher than critical current density $J_c$, especially on the brims of top and bottom surfaces of the disk. $J_1(\rho, z)$ inside the disk is almost zero. The screen current lasts only very short time. If the cooling of the surface of the disk is ideal, the dissipation heat cannot drive the disk to normal state, and the disk will be still in superconducting state. Fig 1(b) shows the distribution of the trapped magnetic field in a half cross section of the disk with $\rho/a = 0 \sim 1$ and $z/a = 0 \sim 1$.

**B. Current $J_2$ profiles in SBM**

As soon as $\dot{B}_a = 0$ and $B_a = 0$, the surface screening current density $J_1(\rho, z)$ rapidly decreases. Fig. 2 shows the profiles of the current density $J_2(\rho, z)$ calculated by Eq. (9), and corresponding distributions of the trapped magnetic field in a half cross section of the disk with $\rho/a = 0 \sim 1$ and $z/a = 0 \sim 1$ at different moments of the initial stages. With the penetration of the current density into the disk center the trapped magnetic field at the side surface of the disk converts its direction.

After about 144 s, $J_2(\rho, z)$ in whole disk reaches a plateau of $0.6 J_c$. Afterwards the distribution of $J_2(\rho, z)$ keeps this plateau feature, while decreasing with time very slowly. At $t = 10000$ s, $J_2(\rho, z)$ decreases to about $0.5 J_c$. Fig. 3 shows the plateau feature of $J_2(\rho, z)$ at 144 s and 10000 s.

**C. Trapped magnetic field**

Fig. 4 shows the decay of the axial component $B_z$ and radial component $B_\rho$ of the trapped magnetic field at different points of the radius on the top surface of the disk as an approximately logarithmic function of time. The axial component $B_z$ at about $\rho/a \sim 0.955$ is almost zero. The direction converse of the axial component $B_z$ at $\rho/a \sim 0.955$ can be clearly observed.

**D. Effect of sample geometry on the trapped magnetic field**

For fixed aspect ratio $b/a$, large disk radius will increase the field of full penetration $H_p$, then the initial applied magnetic field $H_a (= 0.3 H_p)$ rises without augmenting the gradient of the applied magnetic field. The distribution volume of the current density will also enlarge and finally the trapped magnetic field will go up.

The point, at which the direction of the axial component of the trapped magnetic field reverses, is called the neutral point. The position of the neutral point moves to the center of the top surface with the decrease of the aspect ratio $b/a$. The trapped magnetic flux lines bend seriously for the sample with small aspect ratio $b/a$. When the aspect ratio $b/a$ increases, the trapped magnetic flux lines unbend, finally parallel to the z-axis for infinite cylinder. The radial component of the trapped magnetic field decreases with the increase of the aspect ratio $b/a$, indicating larger demagnetization factor for the sample with smaller aspect ratio $b/a$.

The trapped magnetic field at the center on the top surface of the disk, defined as $B_\rho^0$, has only the axial component, which provides us a good gauge for the investigation of the strength of the trapped magnetic field and the magnetic flux creep. Fig. 5 shows the trapped magnetic field at the center on the top surface of the disk as a function of time for different aspect ratios $b/a$. Parallel lines in the figure indicate that the decay of the trapped magnetic fields is independent of the aspect ratios $b/a$.

**E. Effect of superconducting parameters on the trapped magnetic field**

Now we turn to the effect of superconducting parameters of the superconductor, the critical current density $J_c$ and the flux creep exponent $\sigma$, on the trapped magnetic field. Fig. 6a shows the trapped magnetic field at the center of the top surface $B_\rho^0$ as a function of time with different superconducting critical current densities. It is clear that higher the $J_c$, higher the trapped magnetic field $B_\rho^0$, however, the decay curves of the trapped magnetic field for different $J_c$ almost parallel, indicating that the decay is independent of $J_c$. Fig. 6b shows the trapped magnetic field at the center of the top surface $B_\rho^0$ as a function of time with different flux creep exponents $\sigma$. When the flux creep exponent $\sigma$ increases, the trapped magnetic field $B_\rho^0$ rises and the decay rate decreases. It seems that the decay rate of the trapped magnetic field depends only on the flux creep exponent $\sigma$.

**F. Effect of applied magnetic field on the trapped magnetic field**

Fig. 7 shows the trapped magnetic field at the center of the top surface $B_\rho^0$ as a function of time with different applied magnetic inductions ($B_a/B_p = 0.1, 0.2, \cdots, 1$). When the applied magnetic induction $B_a \geq 0.3 B_p$, all curves superpose together after about 10 s, which indicates the trapped magnetic field at the center of the top surface $B_\rho^0$ is independent of $B_a$. It is easy to understand that the trapped magnetic field $B_\rho^0$ is dominated by the sample volume, critical current density and the flux creep exponent. The result indicates clearly that we cannot obtain higher trapped magnetic field $B_\rho^0$ by increasing the applied magnetic induction. At chosen parameters the applied magnetic induction $B_a = 0.3 B_p$ is enough to magnetize the superconductor.

## IV. CONCLUSIONS

According to the equation of motion for current density inside the superconducting disk, the distribution of the current density $J(\rho, z)$ and the profile of the trapped magnetic field of a superconducting disk magnetized by uniform field are calculated from first principles. The superconductor is treated as the material complying with $\boldsymbol{B} = \mu_0 \boldsymbol{H}$ and the flux creep is described by the voltage-current law $\boldsymbol{E} = \boldsymbol{E}_c (J/J_c)^n$. The sample volume, the critical current density and the flux creep exponent dominate the strength of the trapped magnetic field. The flux creep exponent determines the decay rate of the trapped magnetic field. The aspect ratio $b/a$ of the sample, which is related to the demagnetization factor, influences the shape of the trapped magnetic flux lines. At chosen parameters the applied magnetic induction $B_a = 0.3 B_p$ is high enough to magnetize the superconductor.

## ACKNOWLEDGMENTS

This work was supported by the National Science Foundation of China (NSFC 10174004) and the Ministry of Science and Technology of China (Project No. NKBRSF-G1999064602).

Figure captions:

Fig. 1: (a) Profiles of the surface screening current density $J_1(\rho,z)$ calculated by Eq. (8), and (b) corresponding distribution of the trapped magnetic field. The vector length and direction represent the strength and the direction of the trapped magnetic field at the origin point of the vector, respectively.

Fig. 2: Profiles of the current density $J_2(\rho,z)$ calculated by Eq. (9), and corresponding distributions of the trapped magnetic field at different moments in the initial stages. The vector length and direction represent the strength and the direction of the trapped magnetic field at the origin point of the vector, respectively.

Fig. 3: Profiles of the current density $J_2(\rho,z)$ calculated by Eq. (9) at different moments in the initial 144 s and 10000 s. The distribution of $J_2(\rho,z)$ keeps the plateau feature.

Fig. 4: (a) Decay of the axial component $B_z$ and (b) radial component $B_\rho$ of the trapped magnetic field at different points of the radius on the top surface of the disk as an approximately logarithmic function of time.

Fig. 5: Trapped magnetic field at the center on the top surface $B_\rho^0$ as a function of time for different aspect ratios $b/a$.

Fig. 6: (a) Trapped magnetic field at the center of the top surface $B_\rho^0$ as a function of time with different critical current densities $J_c$, (b) or different flux creep exponents $\sigma$.

Fig. 7: Trapped magnetic field at the center of the top surface $B_\rho^0$ as a function of time with different applied magnetic inductions ($B_a/B_p = 0.1, 0.2, \cdots, 1$). When $B_a \geq 0.3 B_p$, all curves superpose together.

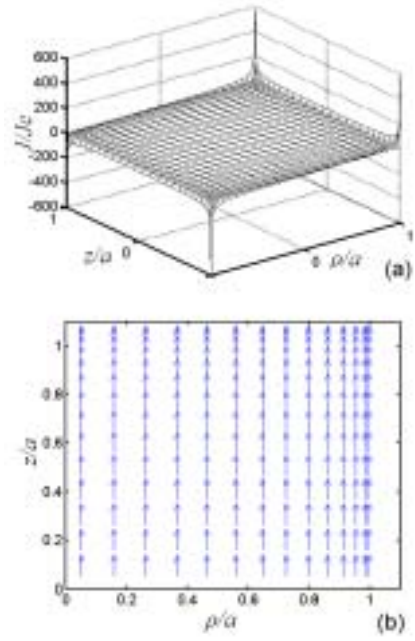

Fig. 1

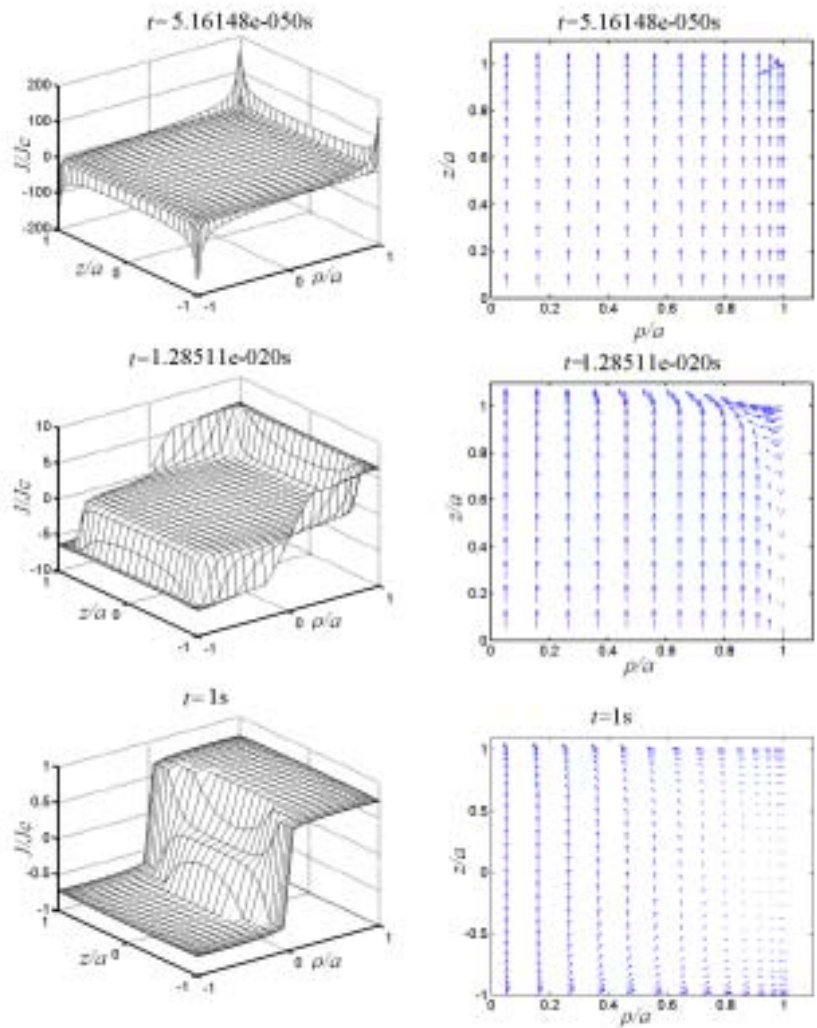

Fig.2

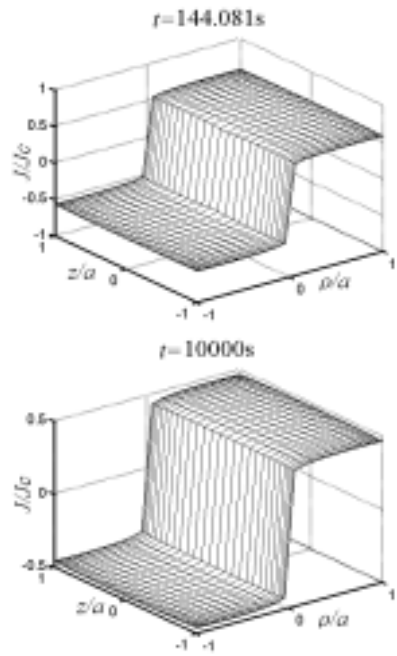

Fig. 3

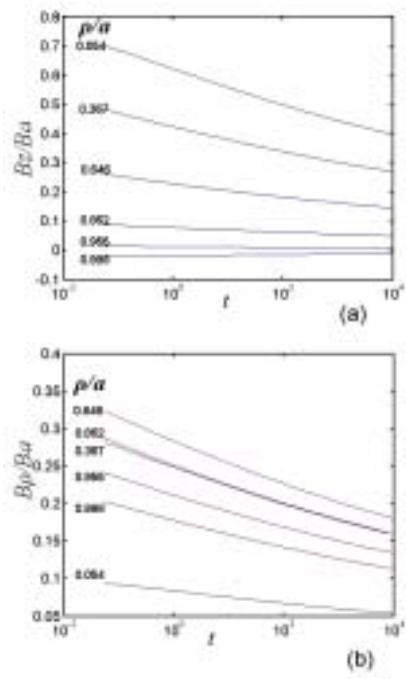

Fig. 4

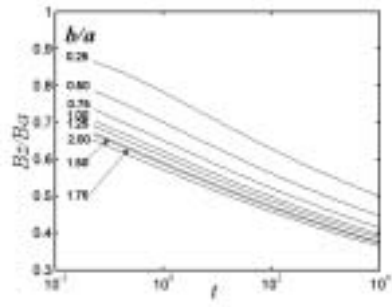

Fig. 5

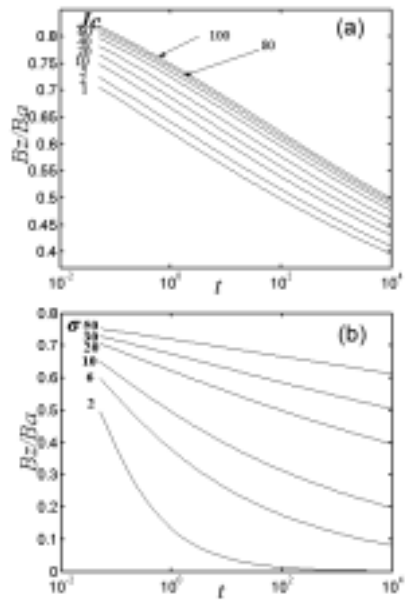

Fig. 6

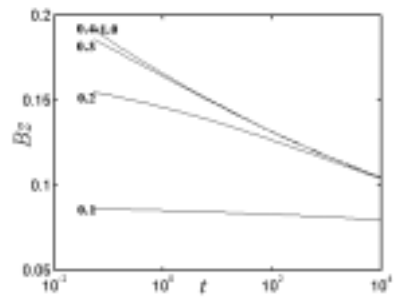

Fig. 7